\documentclass[aps,pre,twocolumn,groupedaddress,floatfix,superscriptaddress]{revtex4}
\usepackage{dcolumn}
\usepackage{amsmath}
\usepackage{amssymb}
\usepackage{lipsum}
\usepackage{graphicx}
\usepackage{bm}
\usepackage[T1]{fontenc}
\usepackage{color}
\usepackage{url}
\usepackage[bf]{subfigure}
\usepackage{rotating}
\usepackage{soul}
\usepackage{scalefnt}
\usepackage{hyperref}
\usepackage{scalefnt}

\usepackage{perpage} 
\MakePerPage{footnote}

\usepackage{color}
\definecolor{redcolor}{rgb}{1.0,0.,0.}

\usepackage[utf8]{inputenc}

\begin{document}
\title{How Integrated are Theoretical and Applied Physics?}

\author{Henrique F. de Arruda}
\affiliation{Institute of Mathematics and Computer Science, University of S\~ao Paulo, S\~ao Carlos, SP, Brazil.}
\author{Cesar H. Comin}
\affiliation{Department of Computer Science, Federal University of S\~ao Carlos - S\~ao Carlos, SP, Brazil}
\author{Luciano da F. Costa}
\email{ldfcosta@gmail.com}
\affiliation{S\~ao Carlos Institute of Physics, University of S\~ao Paulo, S\~ao Carlos, SP, Brazil}

\begin{abstract}
How well integrated are more theoretically and application oriented works in Physics currently?   This interesting question, which has several relevant implications, has been approached mostly in a more subjective way.  Recent concepts and methods from network science are used in the current work in order to develop a more principled, quantitative and objective approach to gauging the integration and centrality of more theoretical/applied journals within the APS journals database, represented as a directed and undirected citation network.   The results suggest a surprising level of integration between more theoretically and application oriented journals, which are also characterized by remarkably similar centralities in the network.
\end{abstract}
\maketitle

\setcounter{secnumdepth}{1}

\section{Introduction}
Science has unfolded on the interface between theory and applications.   For instance, in classical Greece, more considerable attention was given to theoretical constructs, such as Plato's \emph{Theory of Forms}~\cite{fine1993ideas}.  Contrariwise, Roman antiquity concentrated on applications and engineering~\cite{hill2013history}.  Ever since, science progressed in both theory and applications which, as we believe, are equally important and complement one another.  From time to time questions arise regarding the current level of integration between this two orientations~\cite{brooks1994relationship,meyer2010can}. Such questions, which are perhaps implied by the own division of Physics into theoretical and applied branches, are relevant because they have several implications~\cite{brooks1994relationship}.  For instance, this could involve in important theoretical results not reaching the application and technological level, or vice-versa.  Though much of the discussion about the integration between theoretical and applied science has been conducted in a relatively informal and subjective way, we now have the means for approaching it in a more principled, quantitative and, therefore, more scientific way. 

A common approach for quantifying the integration between theory and application is to analyze citation patterns between scientific publications and patents~\cite{tussen2000technological, shibata2010extracting}. Such an analysis can focus on how science influences technology~\cite{meyer2010can, shibata2010extracting} as well as on the benefits that technological advancements can bring to science ~\cite{gazis1979influence}. 

The compilation and dissemination of statistics about publications, such as the APS database, as well as other initiatives~\cite{APS2017}, can be efficiently mapped into intuitive representations, such as graphs or networks~\cite{costa2007characterization, costa2011analyzing}. These representations are capable of making explicit the connectivity and relationship between different scientific areas.  Powerful methods have also become available that can be used to analyze these representations efficiently.  The coming of age of these research possibilities is now directly reflected in the area of \emph{Scientometry}~\cite{bonilla2008scientometric,larowe2008scholarly, amancio2012using}.

\begin{figure}[!ht]
 \centering
 \subfigure[]{\includegraphics[width=0.7 \linewidth]{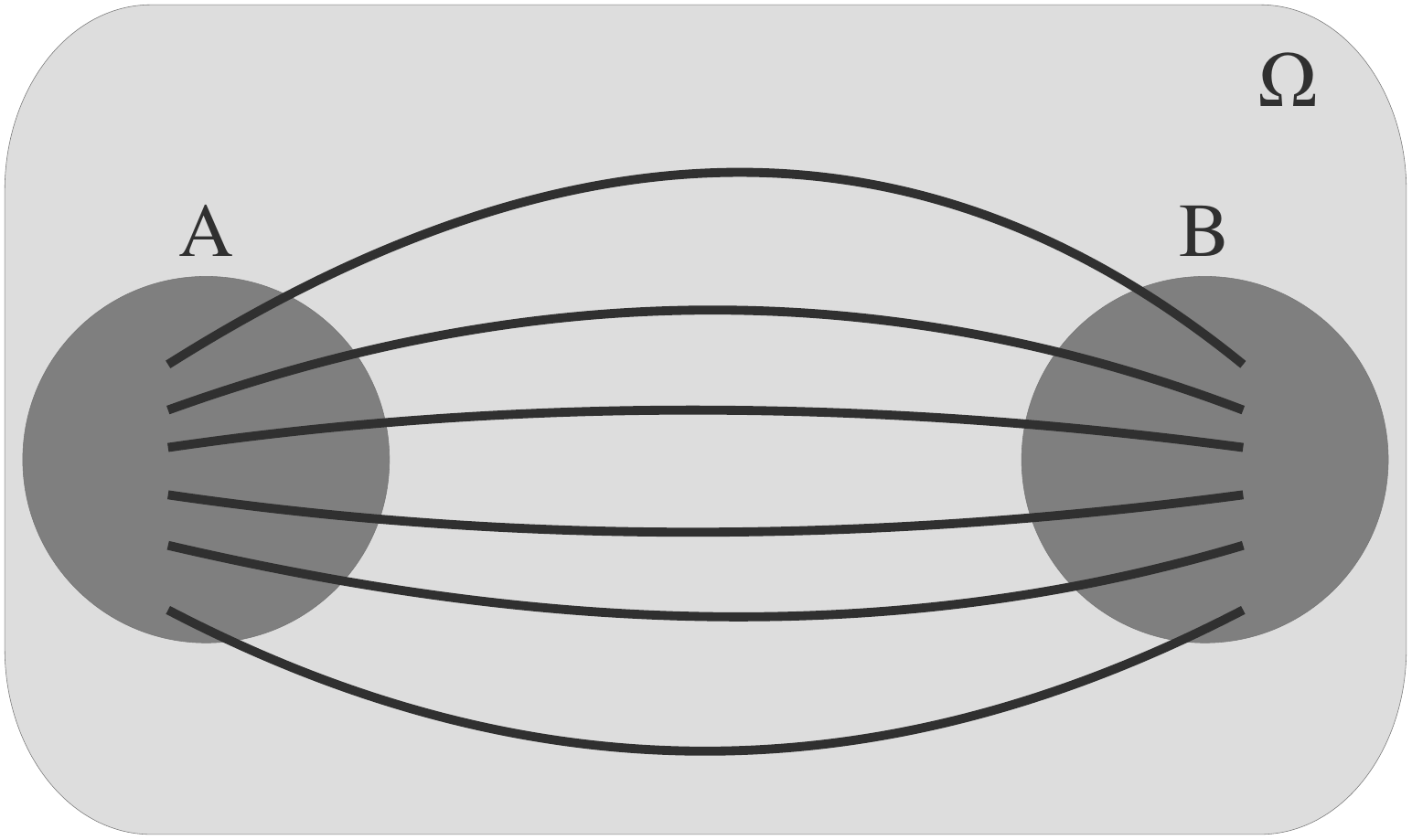}}
 \subfigure[]{\includegraphics[width=0.7 \linewidth]{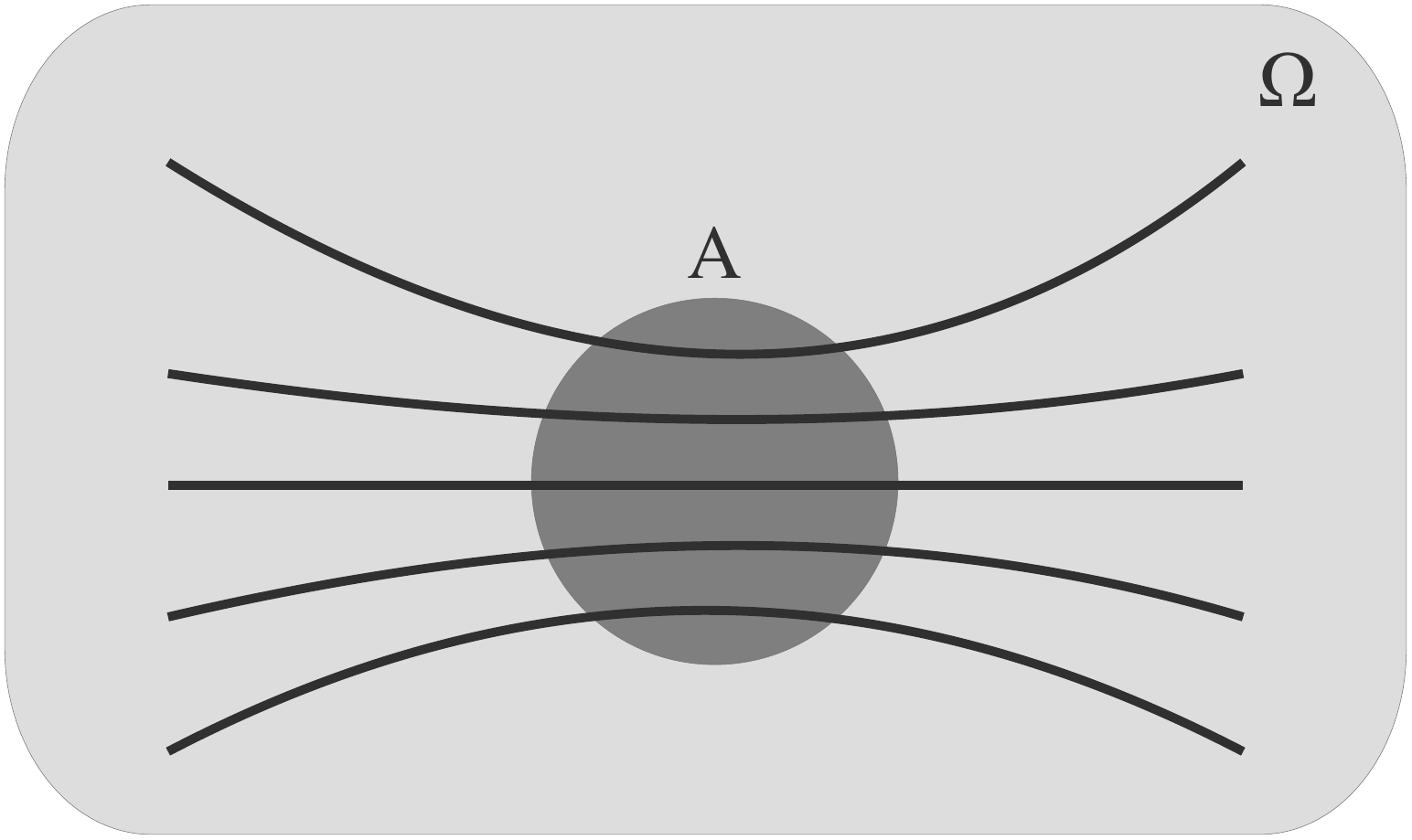}}
 \caption{The integration between two journals A and B can be quantified in terms of the multiplicity of shortest paths between them (a).  The integration of a given journal A with the remainder journals can be further quantified by considering the betweenness centrality of its constituent nodes (b). }
 \label{fig:scheme}
\end{figure}

The current work reports precisely an attempt at investigating the integration between applied and theoretical science, especially from the perspective of Physics.  More specifically, we consider the APS database~\cite{APS2017}, containing several journals with diverse focuses. A single network is constructed from this database, where each node represent an article (all available articles from all journals in the database), while the links stand for citations between these articles.  Then, two specific journals were selected as presenting dominant focus on theory (Phys. Rev. D) and applications (Phys. Rev. Applied), according to the description of their respective scopes.  Because the latter journal is very recent, and therefore includes less publications, we defined a second experiment in which Phys. Rev. Applied was substituted by \emph{Phys. Rev. B}, which is older and has many more articles.  The integration or proximity between the different journals was quantified in terms of the multiplicity of shortest paths between their respective body of articles (Figure~\ref{fig:scheme}a), as well as by the betweenness centrality~\cite{freeman1977set} (Figure~\ref{fig:scheme}b) of the considered groups of articles (Phys. Rev. D, Phys. Rev. Applied, and control groups).  The basic hypothesis here is that the existence of multiple shortest paths of comparable lengths between two sets of articles indicates, to a good extent, integration and relationship between these two sets.  These measurements are complemented by the betweenness of a journal, which is capable of expressing how many of the shortest paths between pairs of articles from the overall database go though that journal.

The results obtained from the aforementioned investigation revealed that journals of more theoretical and applied nature are about as tightly interrelated as any two randomly chosen sets of articles with same size.   These results suggest that theoretical and applied physics are therefore closely related and integrated.  Though these results are specific to the adopted database, network representation and measurements, they nevertheless provide quantitative, objective insights into the relevant question of integration between applied and theoretical physics.  It is hoped that further studies can be developed considering larger databases and other areas, so as to achieve broader and more conclusive results. 

This article starts by presenting the database, network representation and measurements, and then present and discuss the obtained results.   Prospects for further investigations are also highlighted.


\section{Materials and Methods}

In the APS database~\cite{APS2017}, each article is represented by its own Digital Object Identifier (DOI), and the citations are organized as a spreadsheet. For example, if a given article $A$ cites $B$, the pair of DOIs for A and B are incorporated as a row into the table.  Additionally, some meta-data are also included, such as the year of publication and the published journal. We mapped the citation table into a directed network, in which each article represents a node, and the edges correspond to the citations. The directions follow from the citing to the cited articles.  Though this type of directed representation of the citations is inherently more related to the idea of scientific citation, we also consider \emph{undirected} versions of these networks.  The motivation for doing so is that when one article cites another, it is somehow implied that not only the citing article shares elements with the cited work, but that the reciprocal also takes place.  Indeed, the directionality of citations is purely an implication of temporal causality, in the sense that only the later work can cite the earlier article.  

Starting with this network, we divided the nodes (articles) into three groups: \emph{applied physics}, \emph{theoretical physics}, and \emph{control}. The first two groups correspond to Phys. Rev. Applied (alternatively Phys. Rev. B) and Phys. Rev. D, respectively, and articles from the remaining journals were used to define control groups. The theoretical physics area was represented by the journal \emph{Phys. Rev. D}, as this journal includes more academic-related areas, such as field theory, and quantum field theory.  The applied physics articles were obtained from \emph{Phys. Rev. Applied} because this journal typically publishes articles on applications, including engineering and technology. Because this journal is recent, and therefore includes a relatively small number of articles, we also considered a second experiment where this journal was substituted by \emph{Phys. Rev. B}, which is arguably the next more application-oriented journal in the APS database.  These two experiments are henceforth referred to as \emph{Case Study \#1} and  \emph{Case Study \#2}.  Observe that this choice is particular, and other possibilities could be taken into account, including alternative databases.  Though the results reported in this article are specific to these choices, we believe them to be a reasonable indication of what could be found for other choices.  An overview of the characteristics of the areas is shown in Table~\ref{tab:charac}. Note that the number of articles varies for each group, with the applied group being smaller as a consequence of its more recent introduction. 

\begin{table}[]
\centering
\caption{The adopted groups of articles and some of their properties. The used dataset comprises articles from 1893 to 2016. Control~(A) and Control~(B) represent the Control group of articles when, respectively, Phys. Rev. Applied and Phys. Rev. B is employed.}
\label{tab:charac}
\begin{tabular}{|l | c | r|}
\hline	
Group                & Number of articles & Years \\ 
\hline	
Applied (Rev. App.)  &     703      &   2014--2016  \\
Applied (Rev. B)     &   176,381    &   1970--2016  \\
Theoretical (Rev. D) &   80,427     &   1970--2016  \\
Control (A)          &   509,325    &   1893--2016  \\
Control (B)          &   333,647    &   1893--2016 
\\ \hline	
\end{tabular}
\end{table}

In order to gauge the distance between two different areas of study, we employed the distribution of minimal paths, which were computed using the \emph{Dijkstra} algorithm~\cite{dijkstra1959note}. Sets of nodes with the same size were randomly selected for the three groups, and the shortest path distances were calculated. The number of selected articles was chosen as 705, which is approximately the number of items in the smallest group. Note that we did not include in our statistics the pairs of nodes that do not have paths connecting them. We used the measurement $1/($Path length$)$, henceforth called \emph{cohesion}, instead of the shortest paths so as to have a measurement that is directly related to the relationship between articles (the shortest path length would be \emph{inversely} related, being less intuitive). 

As an additional measurement, we considered the betweenness centrality measurement~\cite{freeman1977set}, which quantifies the centrality of network nodes according to the number of shortest paths to which they belong. This measurement is computed as follows
\begin{equation}
B_{k} = \sum_{ij} \frac{\sigma^{k}_{ij}}{\sigma_{ij}} ,
\end{equation}
where $\sigma^{k}_{ij}$ is the number of shortest paths that connect the nodes $i$ and $j$ and crosses $k$, and $\sigma_{ij}$ is the total number of shortest paths connecting $i$ and $j$.

\section{Results and Discussion}

The obtained results for directed and undirected versions of the article networks are reported and discussed respectively in the following two sections.

\subsection{Case Study \#1 - Phys. Rev. Applied}

\begin{figure}
\centering
\includegraphics[width=0.48\textwidth]{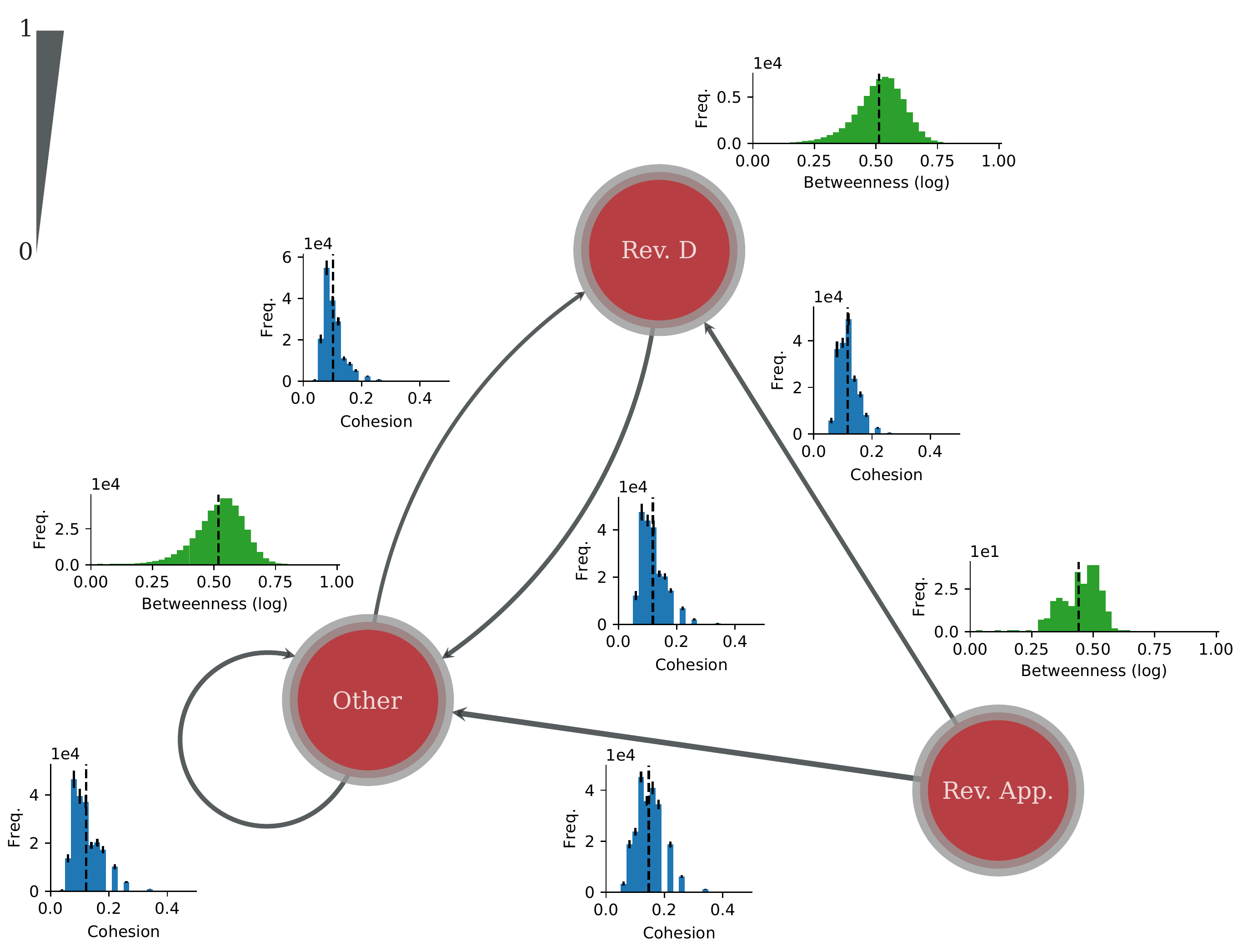}
\caption{\label{fig:directed} Cohesion (blue), defined as the reciprocal of the shortest path length, and betweenness centrality (green) relationships between Phys. Rev. D, Phys. Rev. Applied and Other APS journals for the directed network. The legend at the upper right corner indicates the width range of the links.  Observe that all links have similar width, suggesting symmetric relationships between all groups.  }
\end{figure}

Figure~\ref{fig:directed} shows the results obtained for the directed version of the considered networks.   For the sake of better integration and interpretation of the results, we combined the cohesion histograms with the digraph representing the relationships between the three considered cases, which include: Phys. Rev. D, Phys. Rev. Applied, and the other journals.  The width of each link reflects the average of the respective histogram.  The average values for each histogram are also identified.  The histograms of betweenness centralities obtained for each of the three groups are shown in green.

Though, in principle, all directed interconnections could be expected to appear in this digraph, some of these connections that were too small (e.g.~ between \emph{other} and \emph{applied}) have been discarded.  Generally, there is little difference between the width of the links in the obtained digraph, which indicates a good overall uniformity of connections between the three considered groups.  This result suggests a good integration between the more theoretical and applied articles in Physics.

Regarding the betweenness centrality of the nodes, which is also shown in  Figure~\ref{fig:directed}, similar results were again obtained, with the \emph{Applied} group having a density less symmetric than the other cases, which is a possible consequence of the smaller size of this group.   In addition, the average betweenness centrality observed for this group is slightly smaller than those of the other two groups.

\begin{figure}
\centering
\includegraphics[width=0.48\textwidth]{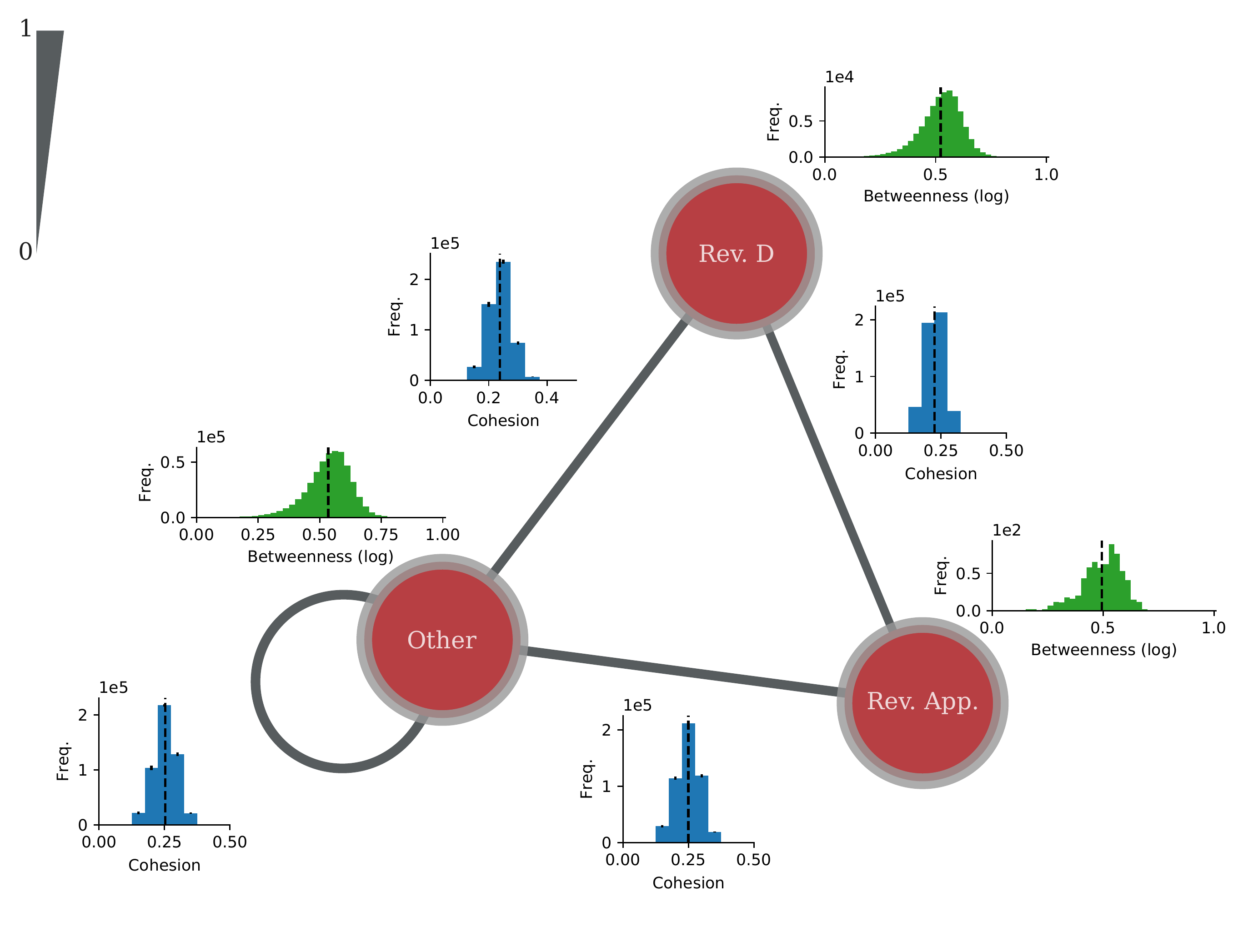}
\caption{\label{fig:undirected} Cohesion (blue) and betweenness centrality (green) relationships between Phys. Rev. D, Phys. Rev. Applied and Other APS journals for the undirected network. The legend at the upper right corner indicates the width range of the links.}
\end{figure}

The results for the undirected versions of the considered networks are presented in Figure~\ref{fig:undirected}, together with the respective cohesion histograms.  The interconnections between the three groups resulted again very similar. 

\subsection{Case Study \#2 - Phys. Rev. B}

\begin{figure}
\centering
\includegraphics[width=0.48\textwidth]{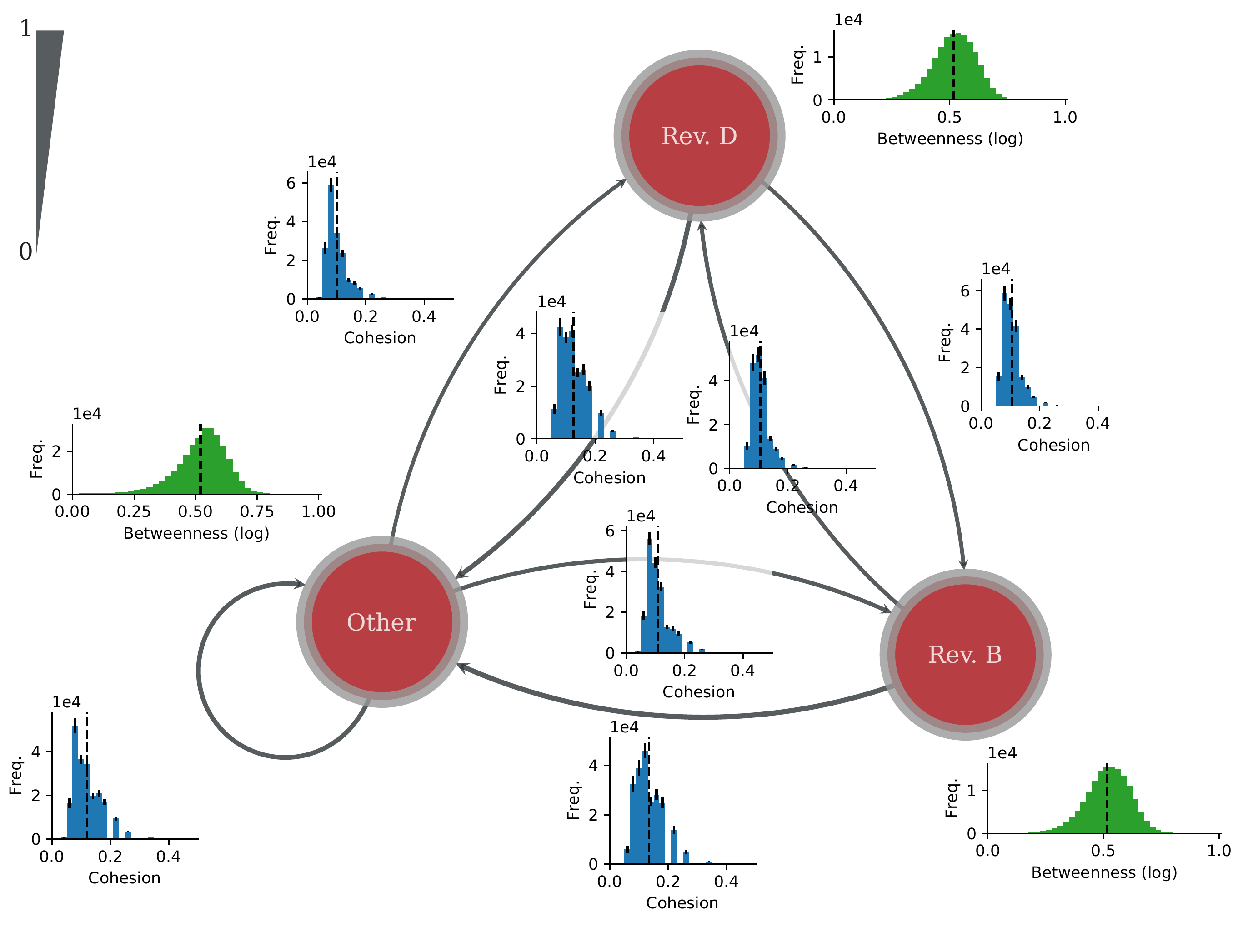}
\caption{\label{fig:directedB} Cohesion (blue) and betweenness centrality (green) relationships between Phys. Rev. D, Phys. Rev. B and Other APS journals for the directed network. The legend at the upper right corner indicates the width range of the links.}
\end{figure}

Figure~\ref{fig:directedB} shows the results for the directed version of the APS network. For this case study, the three considered groups are: Phys. Rev. B, Phys. Rev. D, and the remaining journals.  As before, the width of each link reflects the average of the respective histogram.  Observe that, respectively to the previous results, more links appear in this figure as a consequence of the larger number of articles in Phys. Rev. B.  The cohesion histograms are similar for all cases, indicating that citations between journals have no preferred direction. The betweenness histograms for the groups are also similar. These results are in agreement with those obtained for case study \#1, therefore conforming a good integration between theoretical and applied articles.

\begin{figure}
\centering
\includegraphics[width=0.48\textwidth]{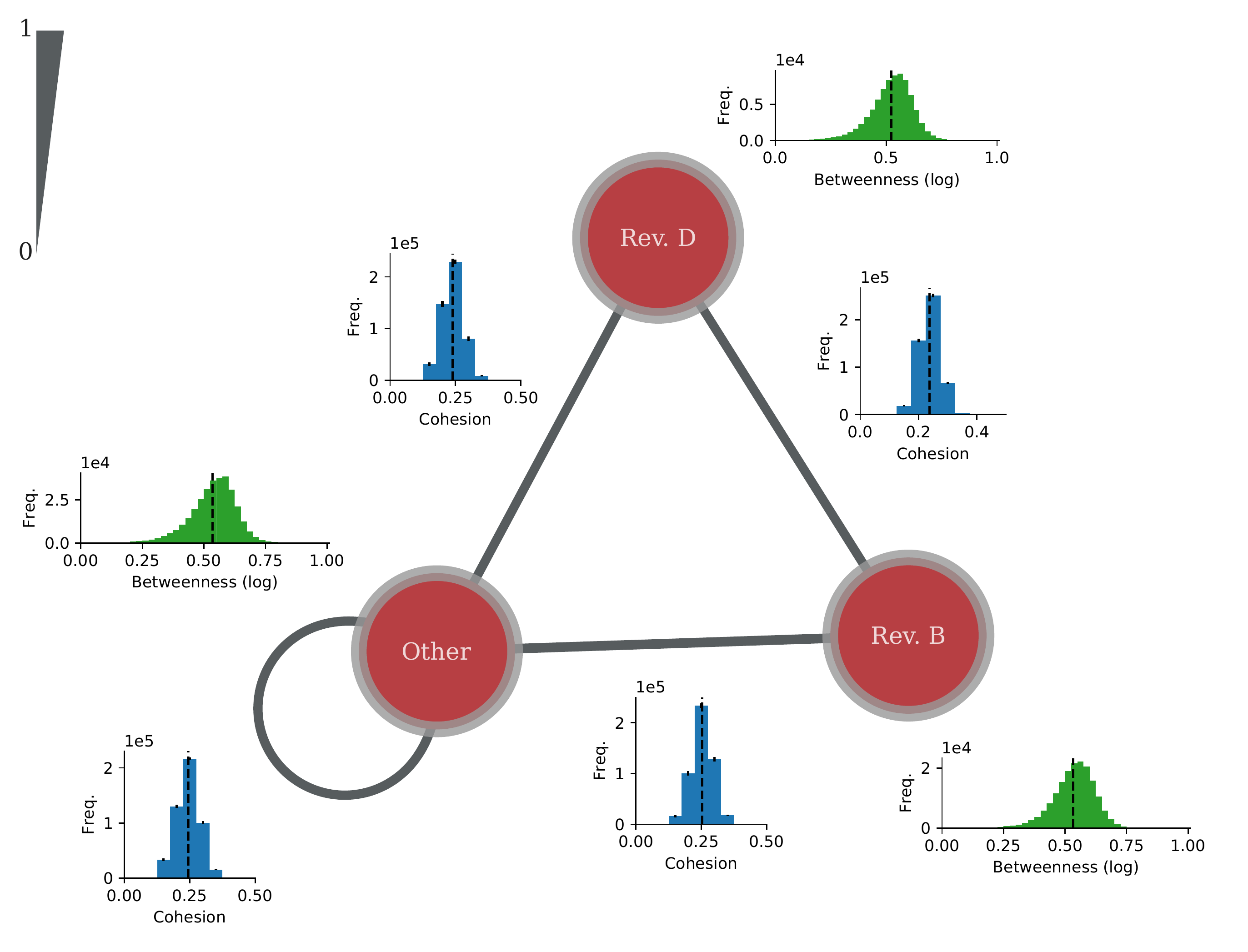}
\caption{\label{fig:undirectedB} Cohesion (blue) and betweenness centrality (green) relationships between Phys. Rev. D, Phys. Rev. Applied and Other APS journals for the directed network. The legend at the upper right corner indicates the width range of the links.}
\end{figure}

The results for the undirected version of the APS network is shown in Figure~\ref{fig:undirectedB}. Again, the cohesion and betweenness histograms are similar in all considered groups.

\section{Conclusion}
Since the beginnings of science, eventual inclinations toward theoretical 
or applied approaches have been subject of considerable debate.  Such discussions are, in principle, justified by the desire to understand how science changes along time, and also because such eventual biases may have effects in the relationship between science and technology.  Though much of these discussions have taken place in a more subjective way, the advent of network science paved the way to more principled, quantitative, and therefore objective approaches to the aforementioned problem.  The current work reports a related attempt.   More specifically, we applied the network science concepts of shortest path and betweenness centrality to quantify interrelationships between journals in networks derived from the interesting APS journal database.  We considered pairwise relations involving potentially more theoretical (Phys. Rev. D) and applied (Phys. Rev. Applied and Phys. Rev. B) APS journals.  

The obtained results indicates a substantial symmetry of relationship and centrality of all considered journals, suggesting that both the more theoretically and applied inclined journals tend to have similar roles and be similarly integrated within the body of APS publications.  So, there seems, at least along the years covered in the APS database, a surprising integration and relationship between more theoretical and applied studies and results in the area of Physics.  We expect that this finding extends potentially to other publications and areas.  Further investigation is however required considering other databases, other types of relationships, other types of measurements, among other possibilities.   It would also be interesting to verify how these results extend to other areas and even science as a whole.  The proposed methodology is also not limited to theoretical/applied orientations of a given area, as it can also be applied to quantify the interrelationship and integration between distinct areas such as Physics and Biology, etc.

\acknowledgments
Henrique F. de Arruda acknowledges CAPES for sponsorship. Cesar H. Comin thanks FAPESP (grant no. 15/18942-8) for financial support. Luciano da F. Costa thanks CNPq (grant no. 307333/2013-2) and NAP-PRP-USP for sponsorship. This work has been supported also by FAPESP grants 11/50761-2 and 2015/22308-2.






\bibliography{references}

\end{document}